\documentclass[review,3p,12pt]{elsarticle}
\usepackage{physics}
\usepackage{amssymb}
\usepackage{amsmath}
\usepackage{bm}
\usepackage{xcolor,soul}
\sethlcolor{white}
\usepackage{xr-hyper}
\usepackage[hidelinks]{hyperref}
\hypersetup{
    colorlinks=False,
    linkcolor=blue,
    filecolor=blue,      
    urlcolor=black,
    pdftitle={Overleaf Example},
    }
\usepackage[version=3]{mhchem} 

\newcommand\rev{\hl}

\journal{Electrochimica Acta}

\begin{document}

\begin{frontmatter}



\title{The effect of buoyancy driven convection on the growth and dissolution of bubbles on electrodes}


\author[inst1]{Farzan Sepahi}
\ead{f.sepahi@utwente.nl}
\author[inst1,inst2]{Nakul Pande}
\author[inst1]{Kai Leong Chong}
\author[inst2]{Guido Mul}
\author[inst1,inst4,inst5]{Roberto Verzicco}
\author[inst1,inst3]{Detlef Lohse}
\ead{d.lohse@utwente.nl}
\author[inst2]{Bastian T. Mei}
\ead{b.t.mei@utwente.nl}
\author[inst1]{Dominik Krug\corref{cor1}}
\ead{d.j.krug@utwente.nl}
\cortext[cor1]{Corresponding author}

\affiliation[inst1]{organization={Physics of Fluids Group, University of Twente},
            addressline={P.O. Box 217}, 
            postcode={7500 AE}, 
            city={Enschede},
            country={Netherlands}}
            
\affiliation[inst2]{organization={Photo Catalytic Synthesis, MESA+ Institute for Nanotechnology, University of Twente},
            addressline={P.O. Box 217}, 
            postcode={7500 AE}, 
            city={Enschede},
            country={Netherlands}}
            
\affiliation[inst3]{organization={Max Planck Institute for Dynamics and Self-Organization},
            addressline={AM Fassberg 17}, 
            postcode={37077}, 
            city={Göttingen},
            country={Germany}}
            
\affiliation[inst4]{organization={Dipartimento di Ingegneria Industriale, University of Rome ‘Tor Vergata’},
            addressline={Via del Politecnico}, 
            postcode={37077}, 
            city={Roma 00133},
            country={Italy}}
            
\affiliation[inst5]{organization={Gran Sasso Science Institute},
            addressline={Viale F. Crispi}, 
            postcode={7}, 
            city={6700 L’Aquila},
            country={Italy}}

\begin{abstract}
		Enhancing the efficiency of water electrolysis, which can be severely impacted by the nucleation and growth of bubbles, is key in the energy transition. In this combined experimental and numerical study, in-situ bubble evolution and dissolution processes are imaged and compared to numerical simulations employing the immersed boundary method.
		We find that it is crucial to include solutal driven natural convection in order to represent the experimentally observed bubble behaviour even though such effects have commonly been neglected in modelling efforts so far. We reveal how the convective patterns depend on current densities and bubble spacings, leading to distinctively different bubble growth and shrinkage dynamics. Bubbles are seen to promote the convective instability if their spacing is large ($\geq4$mm for the present conditions), whereas the onset of convection is delayed if the inter-bubble distance is smaller. Our approach and our results can help devise efficient mass transfer solutions for gas evolving electrodes.
\end{abstract}




\end{frontmatter}


	\section{Introduction}
	
	The process of bubble formation is of significant technological relevance \cite{Lohse2018}. This also holds in the context of industrial processes relevant for the energy transition such as water electrolysis or electrochemical $\ce{CO2}$ reduction \cite{Zeng2010,Angulo2020,Sacco2019}. Production of `green' hydrogen from water splitting is envisioned to be a major contributor in the future energy mix \cite{IEA2019}. However, current technologies suffer from limited cell efficiencies or high costs \cite{IEA2019,DeGroot2021}, rendering large scale operation uneconomical in many cases. It is well established that the presence of bubbles critically affects electrolyser efficiency \cite{Angulo2020,Zhao2019,DeGroot2021}, e.g by reducing the active electrode area \cite{Eigeldinger2000,Vogt2005} or by raising the cell resistance \cite{Sides1980,Dukovic1987}. This has sparked significant interest in concepts to manage the bubble nucleation and growth and the gas flow on gas-evolving electrodes \cite{Kadyk2016,Brussieux2011,Xu2018,Penas2019,Pande2019}. For such approaches, it is crucial to understand the mass transport phenomena, as they determine the bubble nucleation, growth and detachment rates \cite{Oguz1993,Yang2015,VanDerLinde2017}.
	
	With the exception of recent work on local Marangoni convection \cite{Yang2018,Massing2019,Hossain2020,Meulenbroek2021}, related studies are mostly performed assuming a stagnant electrolyte and focus on diffusive transport \cite{Penas2019,VanDerLinde2017,Kadyk2016,Yang2015}. At the same time, the relevance of global convective instabilities in electrochemical systems is now well documented. These can originate from electric fields \cite{Mani2020,Pande2021}, but predominantly also from buoyancy forces resulting from the density gradients caused by electrode reactions and ion transport \cite{Novev2018,Obata2020,Babu2019,Ngamchuea2015}.
	In particular, the simulations of \citeauthor{Ngamchuea2015} \cite{Ngamchuea2015} showed that such solute driven natural convection can significantly enhance mass transport during the oxidation of hexacyanoferrate, while later studies also accounted for thermal forcing \cite{Novev2016,Novev2017}. 
	The presence of natural convection in water electrolysis has also been demonstrated experimentally indirectly through pH-mapping \cite{Obata2020} and directly through velocity measurements \cite{Babu2019}. 
	
	The presence of convection over a wide parameter range strongly suggests that this effect also plays a role in the bubble evolution. This is corroborated by the fact that e.g. \citeauthor{VanDerLinde2017} \cite{VanDerLinde2017} had to scale the actual current densities down by a factor of up to 10 in order to match experimentally measured electrolytic bubble growth rates, as models assuming pure diffusion strongly overpredicted the bubble growth. Given such inconsistencies, it is our goal here to systematically explore the role of convective effects on the bubble evolution in electrochemical water splitting. Moreover, this work provides insight into how the presence of bubbles in turn affects the hydrodynamic instability. Our approach combines experiments with direct numerical simulations (DNS) employing the immersed boundary method. Details on both will be provided in the next section before we will present and discuss the results and summarize our findings in the conclusion.
	
	\begin{figure*}
		\centering
		\includegraphics[scale = 1]{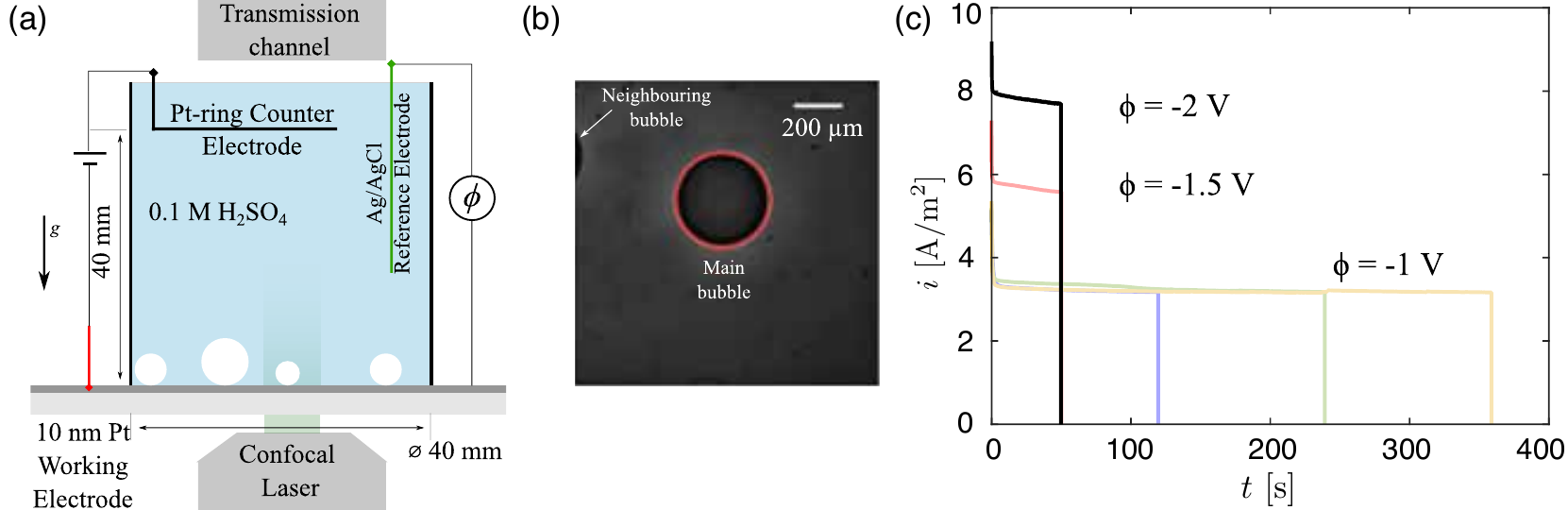}
		\caption{(a) Schematic of the experimental setup. (b) Sample transmission image with the red line indicating the extracted bubble size.
		(c) Measured current densities, $i$, for the different pulses at varying constant potentials ($\phi = -1~\text{V}, -1.5 ~\text{V}, -2~\text{V}$) and pulse times $\tau_p$ (evident from the drop to 0 in $i$).}
		\label{fig:ExpSetup}
	\end{figure*}

    \section{Experimental and numerical details}

	\subsection{Experimental setup}\label{sec:Exp}
	
	The electrochemical cell (see Fig. \ref{fig:ExpSetup}(a)) is made of Teflon and houses a typical undivided 3-electrode configuration: A transparent platinum (Pt) working electrode, a Pt mesh counter electrode shaped as a ring and placed at a distance of $\approx 4~\text{cm}$ from the working electrode, and a Ag/AgCl (in 3M NaCl; BasiR) reference electrode. The setup was mounted on the stage of a Nikon A1R confocal microscope and illuminated from below with a 532 nm laser. Partial transparency of the working electrode was achieved by evaporating 10 nm Pt on glass, with a 3 nm Chromium underlayer (10 nm Pt roughly $\approx~30\%$ transmittance \cite{Heavens1955}). In this way, bubbles appeared as shadows in the transmission images as shown in Fig. \ref{fig:ExpSetup}(b). The cell was operated using a VersaStat (PAR) potentiostat with a sampling rate of 100 Hz. Sulfuric acid (0.1 M $\ce{H2SO4}$,  Sigma Aldrich)) was used as electrolyte. 
	
	Simultaneous electrochemical and optical measurements were performed with the following experimental protocol. First, a negative (reduction) potential pulse was applied for a short time ($60~\text{s}-360~\text{s}$ depending on the experiment). \rev{The pulse length and intensity was chosen such that a limited number of bubbles was nucleated and started to grow on the electrode while avoiding disturbances by bubble detachment.} The current density was recorded (see Fig. \ref{fig:ExpSetup}(c)) and the microscope stage was slowly moved (about the electrode center) until a growing bubble was encountered in the field of view of the camera ($1.28 \times 1.28~\text{mm}^2$). \rev{Hence, the bubble measurements typically only start some time after the start of the current pulse.}
	We ensured that the  measured bubble was the first bubble growing at that location to avoid history effects due to depletion of the gas concentration and bubble detachment \cite{MorenoSoto2017,Penas-Lopez2017}. The microscope imaging was continued for approximately $10$ min after the potential pulse to capture the evolution of the bubble size. The open-circuit potential of the cell was measured simultaneously. Fresh electrolyte was used for each individual experiment.  Note that the bubbles are not isolated as can be seen from Fig. \ref{fig:ExpSetup}(b) (here with center-to-center distance $\approx 0.6$ mm) and that we only track the size of the `main' bubble in the field of view.

	\subsection{Simulations}\label{sec:sim}
	
	\begin{figure}[t!]
		\centering
		\includegraphics[width = 0.8\columnwidth]{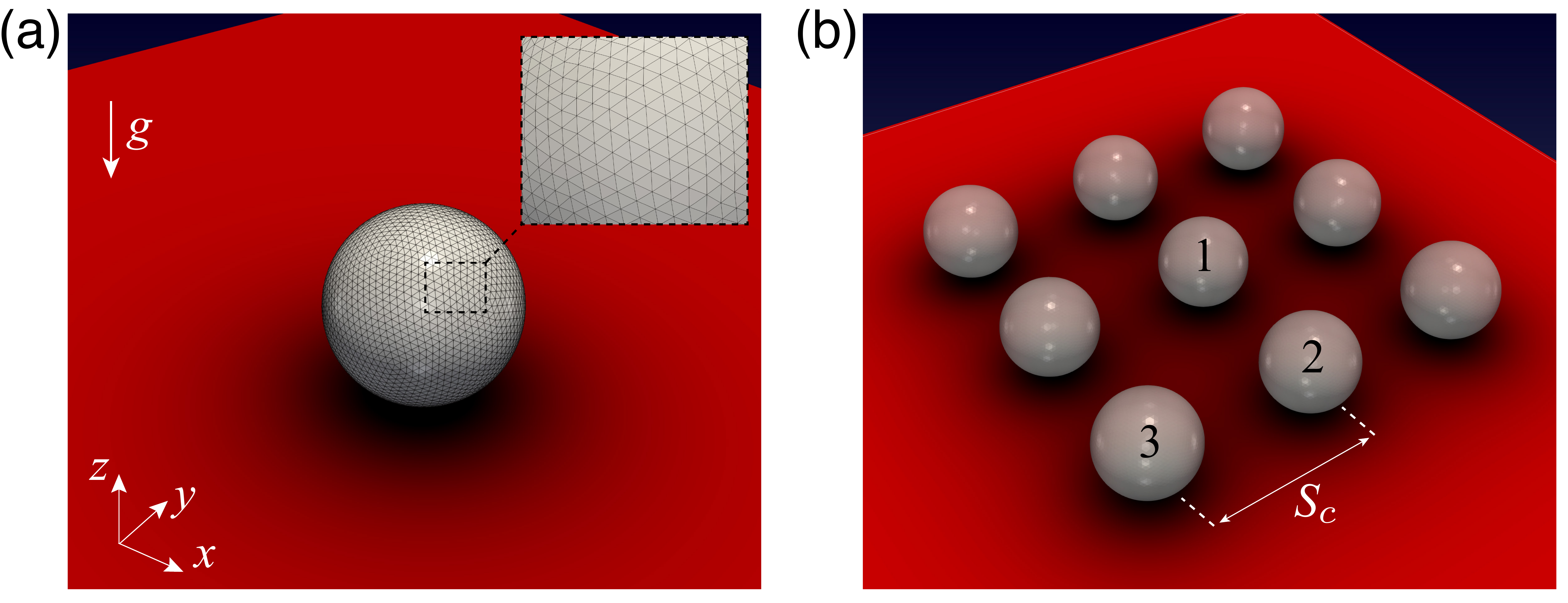}
		\caption{Rendering of (a) the basic simulation setup with a single bubble in the center of the domain and (b) a $3\times3$ bubble cluster with spacing $S_c$.}
		\label{fig:setup}
	\end{figure}
	
    The electrolyte consists of sulfuric acid \rev{which is assumed to fully dissociate in water to hydrogen and sulfate ions as}
	\begin{equation}\label{dissociation}
		\mathrm{H_2SO_4 \rightarrow 2H^+ + SO_4^{2-}},
	\end{equation}
	\rev{which greatly simplifies the numerical modelling.} 
	Additionally, it is assumed that proton reduction to hydrogen is the only cathodic reaction occurring, i.e. 
	\begin{equation}\label{CathodicReaction}
		\mathrm{2H^+ + 2e^- \rightarrow H_2}.
	\end{equation} 
	\rev{Note that given the low current densities employed here, we have neglected the bulk water dissociation reaction for simplicity.}

	To obtain the fluid velocity $\textbf{u}$ field, we solve the Navier-Stokes equations
		\begin{equation}\label{NS}
			\frac{\partial \textbf u}{\partial t} +   \left ( \textbf u \cdot \bm \nabla \right ) \textbf u = -\bm \nabla  p + \nu \bm \nabla^2 \textbf u + \textbf f,
		\end{equation}
		along with continuity,
		\begin{equation}\label{cont}
			\bm \nabla \cdot \textbf u = 0.
		\end{equation}
    Here, $p$ and $\nu$ respectively denote the kinematic pressure and  the kinematic viscosity, and $\textbf{f}$ is the body force due to buoyancy.  Assuming electroneutrality in the bulk of the solution \cite{Dickinson2011} allows us to eliminate the migration terms \cite{Morris1963} (see \hyperref[sec:AA]{Appendix A} for derivation), such that the transport of all species $C_j$ is governed by an effective advection diffusion equation 
    \begin{equation}\label{salt}
			\frac{\partial C_j}{\partial t} +  \left ( \textbf u \cdot \bm \nabla \right ) C_j =  D_j  \bm \nabla^2 C_j.
    \end{equation}
    where the subscript $j=(\ce{s},\ce{H2})$ refers to \ce{H2SO4} and \ce{H2}, respectively.  The diffusivity of $\ce{H2SO4}$ is related to the diffusivity of its ions and is calculated as \cite{Morris1963}:
	\begin{equation}\label{SaltDiff1}
			D_{\ce{s}} = \frac{D_1 D_2  \left( z_1 - z_2 \right)}{z_1 D_1 - z_2 D_2},
		\end{equation}
    where $z_k$ is the ionic valence and subscript $k=(1,2)$ refers to \ce{H+} and \ce{SO4^2-} ions, respectively \rev{and the diffusion constants for the hydrogen and ionic species are given in table {\ref{table1}} in \mbox{\hyperref[sec:AB]{Appendix B}}}.
	
	We employ no slip at the electrode surface and the set of boundary conditions for the scalar fields is (see \hyperref[sec:AA]{Appendix A} for the derivation of Eq. \eqref{cationBC})
	
	\begin{subequations}\label{set2}
		\begin{equation}\label{cationBC}
			\frac{i}{(n_e / s_1)F} = 2D_1 \left(1 - \frac{z_1}{z_2} \right) \left( \frac{\partial C_s}{\partial z} \right) _{z=0},
		\end{equation}
	\begin{equation}\label{gasBC}
			\frac{i}{(n_e / s_{\ce{H2}})F} = D_{\ce{H2}} \left( \frac{\partial C_{\ce{H2}}}{\partial z} \right) _{z=0},
		\end{equation}
	\end{subequations}
	 where $s_j$ and $n_e$ refer to stoichiometric coefficients and the number of transferred electrons in the cathodic reaction \eqref{CathodicReaction}, respectively, and  $F=96 \ 485~\mathrm{C \ mol^{-1}}$ is the Faraday constant. 
	
	Thermal effects are expected to be small in the current system \cite{Vogt1993Temp} and we therefore only consider solutal changes to the density field. Within the Boussinesq approximation of small density changes relative to the initial electrolyte density, the buoyancy force in Eq.  \eqref{NS} is then given by
	\begin{equation}\label{VF}
		\textbf f =   \sum_{\forall \: j} \beta_j \left (  C_j - C_{j,0} \right ) \textbf g,
	\end{equation}
    where $\beta_j$ is the (isothermal and isobaric) volume expansion coefficient of species $j$, $C_{j,0}$ denotes the initial concentration, and $\textbf{g}$ is the gravitational acceleration.
	
    The shape of the bubbles is modelled using an immersed boundary method (IBM), for which specifics are provided in the \hyperref[sec:AB]{Appendix B} along with further details on the numerical method. By evaluating the flux $D_{\ce{H2}} \int_\Sigma \bm{\nabla} C_{\ce{H2}}.\hat{\textbf n} \ d\Sigma$ of $\ce{H2}$ over the bubble surface $\Sigma$ with normal  $\hat {\textbf{n}}$ and using the ideal gas law, we find for the radius $R$ of the (spherical) bubble  
	\begin{equation}\label{radius}
		\frac{dR}{dt} = \frac{{\cal R} T_\infty}{P_0} \frac{1}{4 \pi R^2} \int_\Sigma D_{\ce{H2}}  \bm{\nabla} C_{\ce{H2}}.\hat{\textbf n} \ d\Sigma,
	\end{equation}
    with $\cal R$, $P_0$, and $T_\infty$ denoting the universal gas constant, ambient pressure, and temperature, respectively. \rev{Further, the Laplace pressure is neglected since it is insignificant ($<$1440 Pa while the ambient pressure $p_0=  10^5$ Pa) for the relatively large bubble radii (simulations commence from $R_0=0.1$ mm) considered here. }

    A fixed saturation concentration $C_{\ce{H2},sat}$ is enforced for $\ce{H2}$ at the bubble boundary, while a no flux condition is used for all other species. We further employ a no slip condition at the bubble surface to mimic a fully contaminated bubble \cite{Takagi2011}.
	
    We refrain from modelling the intricacies of the bubble nucleation \cite{Edwards2019,Liu2021}, as this is beyond the scope of the present study. Instead, we initiate bubbles 28 s after the start of the potential pulse with an initial radius $R_0 = 0.1$ mm, which is in accordance with the experiments (see section \hyperref[sec:Exp]{Experimental setup}). Bubbles remain attached tangentially to the electrode surface (contact angle $0^\circ$) throughout the simulations. \rev{This choice well approximates experimental results }\cite{,Janssen1973,Vogt2005} \rev{and conforms with earlier modelling approaches} \cite{Vogt2011,Vogt2015}. In the basic configuration (see Fig. \ref{fig:setup}(a)), we consider a single bubble in the center of the domain and periodic boundary conditions to represent an idealized, regular bubble array with spacing $S$ determined by the lateral dimension of the computational box. \rev{Additionally, we perform simulations in which the single bubble is replaced by a $3\times 3$ array of bubbles with interspacing $S_c$ as shown in Fig. {\ref{fig:setup}}(b) in order to investigate collective effects.}

	\begin{figure*}[t!]
		\includegraphics[scale = 1]{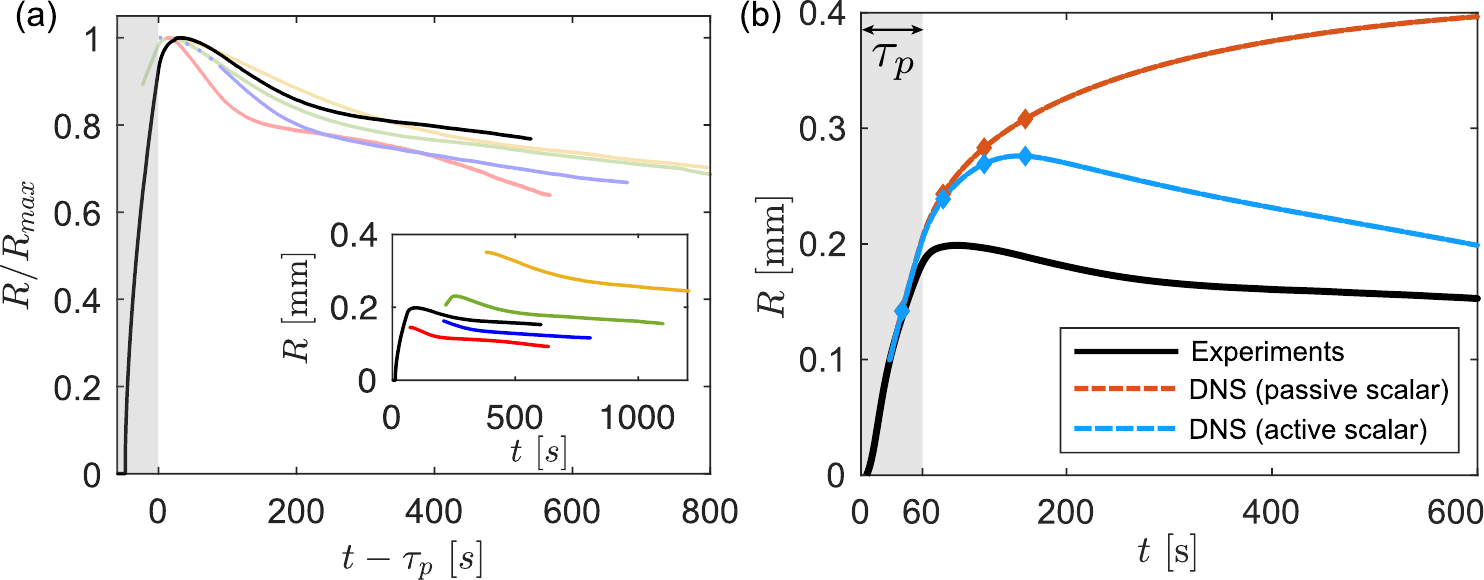}
		\caption{(a) Bubble radius as function of time as obtained from the experiments. The line colour indicates different shapes pulse lengths with the same colour code as in Fig. \ref{fig:ExpSetup}(c). (b) Comparison of experimentally measured bubble radius and those obtained from numerical simulations of a single bubble with (``active scalar'') and without (``passive scalar'') convection being considered.}
		\label{fig:dissolution}
	\end{figure*}

	\section{Results and discussion}
	The inset of Fig. \ref{fig:dissolution}(a) shows the temporal evolution of the bubble radius $R(t)$ for the different potential pulses displayed in Fig. \ref{fig:ExpSetup}(c) (with correspondences indicated by matching line colors). The same data is re-plotted in the main panel of Fig. \ref{fig:dissolution}(a). Shifting the time axis by the respective pulse duration $\tau_p$ and normalizing with the maximum radius $R_{max}$, highlights the similarity of the bubble behaviors in all cases. The most salient feature of this behaviour is the fact that the initial fast bubble growth is followed by a dissolution phase already shortly after the end of the potential pulse. Dissolution is more rapid initially and then reduces to slightly lower rates of dissolution at later times. 
	
	In the following, we will focus on the experiment performed at $\phi = -2$V and $\tau_p =60$ s (black line in Fig. \ref{fig:ExpSetup}(c) and \ref{fig:dissolution}(a)). Here, a bubble happened to nucleate within the initial field of view such that both, the bubble growth and dissolution phases, were captured. In Fig. \ref{fig:dissolution}(b), we compare this bubble evolution to simulation results. In the DNS, we used the experimentally determined current density as an input and chose a box size of $S = 4$ mm, which corresponds to a rough estimate of the typical bubble spacing in the experiments. The importance of convective phenomena is highlighted through a simulation with pure diffusive transport only (setting $\mathbf{f}=0$ in Eq. \eqref{NS}). In that case, the bubble exhibits continued growth even at late times. In contrast, the simulation with active scalars captures the actual bubble behaviour much more faithfully as evidenced by a dissolution phase, i.e. a shrinking of the bubble radius, that sets in shortly ($\approx 100$ s) after the current is stopped. 
	
	\begin{figure*}[t!]
		\includegraphics[ width=1\columnwidth]{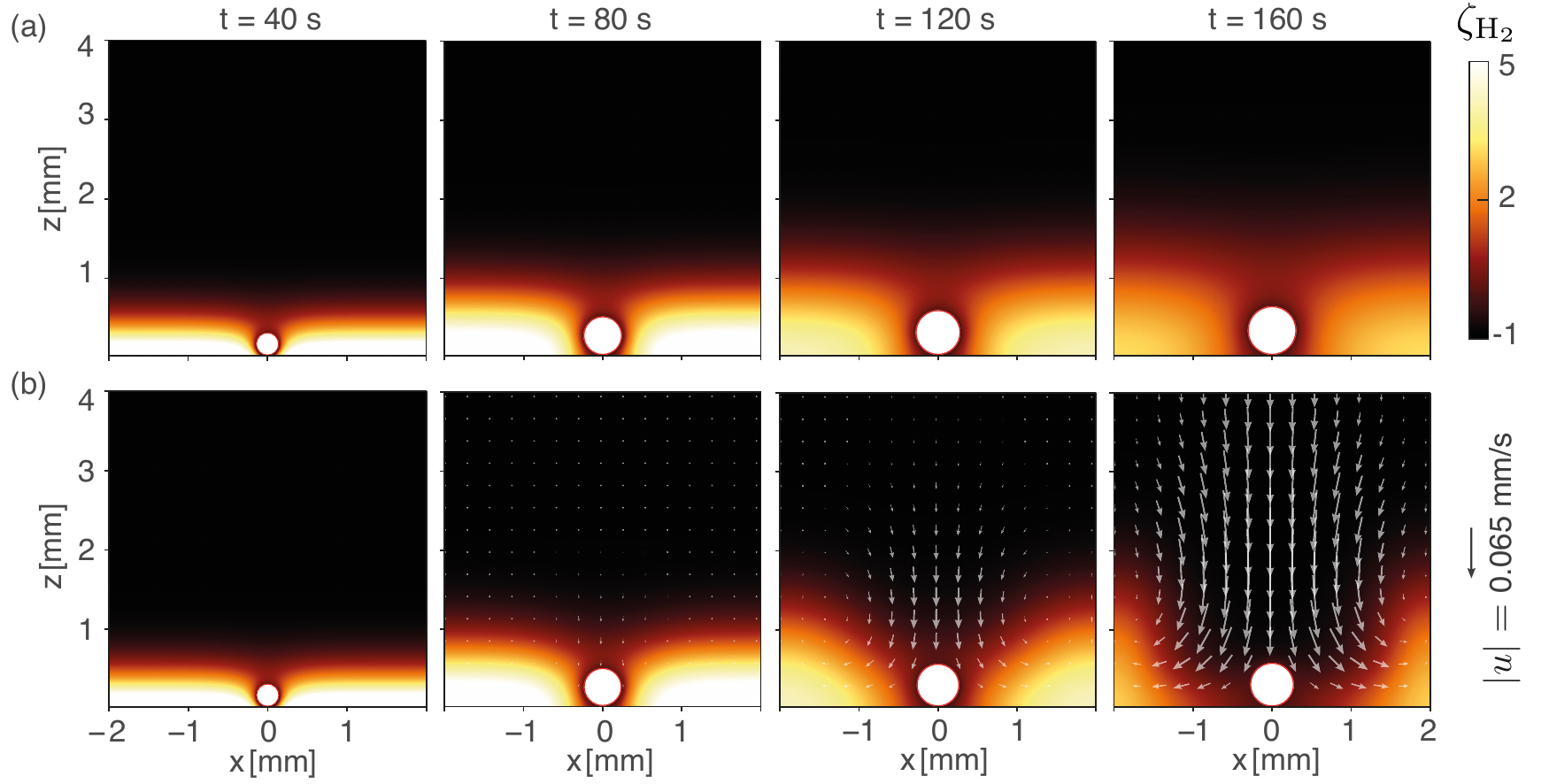}
		\caption{Snapshots of hydrogen supersaturation along with velocity vectors for simulations with passive (a) and active scalars (b). \rev{The reference vector applies to all panels in (b)}. The current density is taken from the experimentally measured values (black curve in Fig. \ref{fig:ExpSetup}(c)). The color code shows the hydrogen oversaturation $\zeta_{\ce{H2}}$. \rev{Full movie is available in the supplementary content.}} 
		\label{fig:ConvectionEffect}
	\end{figure*}
	
	\begin{figure}[h!]
	\centering
		\includegraphics[width=0.5\columnwidth]{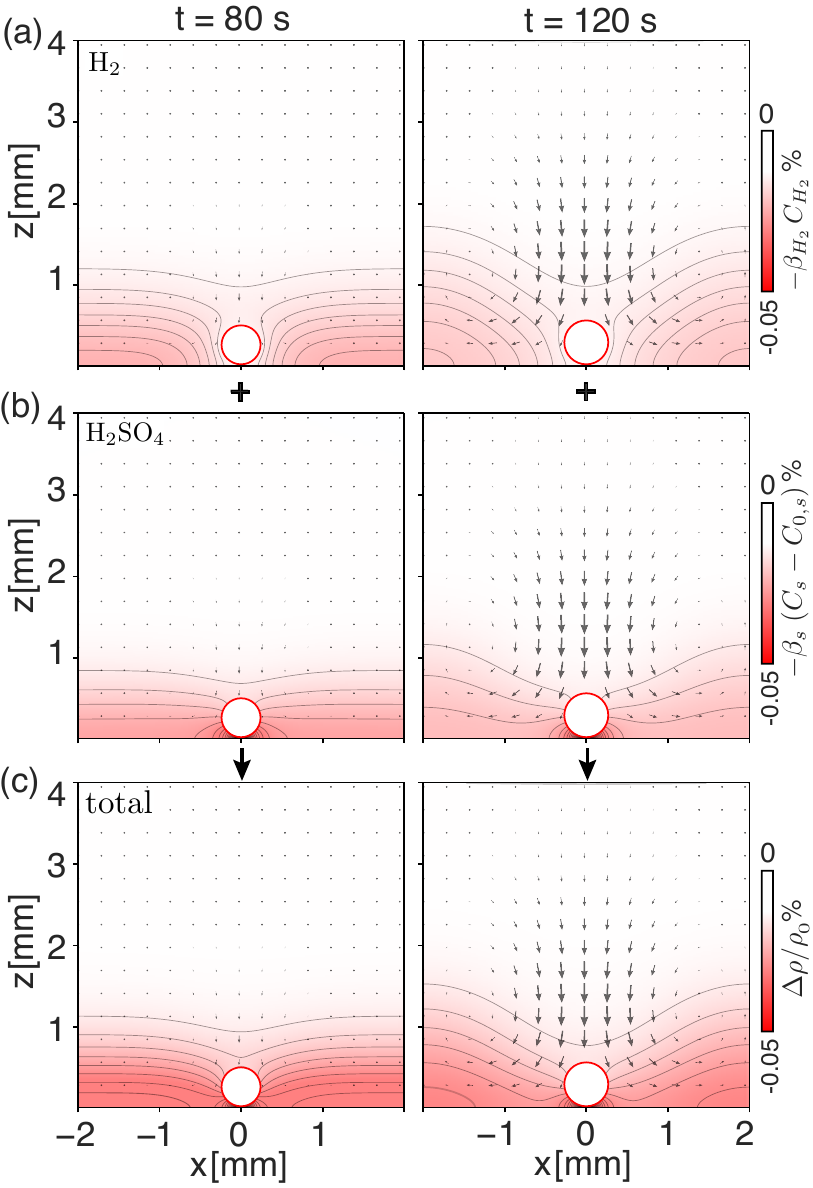}
		\caption{Contribution of local concentration variation of (a) hydrogen and (b) sulfuric acid to the (c) total density fluctuations inside the electrolyte at $t=80$ s (right panels) and $t=120$ s (left panels).}
		\label{fig:DensityContours}
	\end{figure}
	
	The mechanism behind the different behaviour is best illustrated by Fig. \ref{fig:ConvectionEffect}, where the  hydrogen oversaturation  ($\zeta_{\ce{H2}} = C_{\mathrm{H_2}}/C_{\mathrm{H_2,sat}} - 1$) is depicted at several instances in time (indicated as markers in Fig. \ref{fig:dissolution}(b)). 
	Initially, for $t\lessapprox 80$ s the production of $\ce{H2}$ at the electrode leads to a significant local oversaturation, which spreads by pure diffusion. In the case without buoyancy (Fig. \ref{fig:ConvectionEffect}(a)), this also holds at later times. The bubble therefore remains in a boundary layer in which $\zeta_{\ce{H2}} > 0$ even after the potential pulse and therefore continues to grow throughout the entire simulation. The case with buoyancy (Fig. \ref{fig:ConvectionEffect}(b)) starts to differ significantly from this scenario beyond $t \approx 80$ s. This is due to the emergence of a downdraft onto the bubble, which is prominent at $t = 120$ s and even more pronounced at $t=160$ s. The effect of this downflow is to displace the $\ce{H2}$ layer locally, thereby exposing the bubble to undersaturated ( $\zeta_{\ce{H2}} < 0$) electrolyte and leading to its dissolution. 
	
	These observations lead to two relevant conclusions. Most importantly, they show that the experimental findings cannot be explained by considering pure diffusive transport, but are suitably described by including the effects of natural convection. A more subtle point is that the presence of the bubbles and in particular their spacing in turn seems to have an impact on the convective pattern. After all, the position of the plumes relative to the bubbles appears not to be random. The quick dissolution of all experimentally studied bubbles (Fig. \ref{fig:dissolution}(a)) suggests that their location in a downdraft with low gas content is a consistent feature. To investigate how this pinning of the convective pattern to the bubble comes about, we show the distribution of the density change $\Delta \rho$ relative to the background density $\rho_0$ in Fig. \ref{fig:DensityContours}. Variations in $\Delta \rho$ result from the depletion of $\ce{H2SO4}$ as well as from the concentration of $\ce{H2}$. As Fig. \ref{fig:DensityContours} demonstrates, both of these effects act to decrease the local density close to the electrode as a consequence of the reaction there. Further, their contributions are of similar magnitudes for the present conditions.
	However, due to the mass transfer into the bubble, the concentration of $\ce{H2}$ in the vicinity of the bubble is lower, such that the electrolyte density remains somewhat higher there.\footnote{Note that the effect can be opposite for other dissolved gases, e.g. $\ce{CO2}$, for which $\beta >0$, such that depletion causes the local density to decrease \cite{Enriquez2014}.} The presence of the bubble further inhibits the diffusion of the sulfuric acid away from the electrode, which has the same effect on $\Delta \rho$. This results in a lateral density gradient within the concentration boundary layers. The relatively denser fluid around the bubble then favours a downdraft in this region and the emission of lighter electrolyte in the form of plumes in the space between bubbles. 

    It is remarkable that  $\Delta \rho/\rho_0$ remains below $ 0.05 \%$ in the simulations. Yet, consistent with earlier studies \cite{Ngamchuea2015}, this is enough to drive a significant convective flow. We further note that while there is qualitative agreement between experiment and DNS in Fig. \ref{fig:dissolution}(b), quantitative differences remain. We will analyse the reasons for these by exploring the parameter space of varying current densities $i$ and bubble spacings $S$ next.
	
	\begin{figure*}[t!]
	\centering
		\includegraphics[width=1\columnwidth]{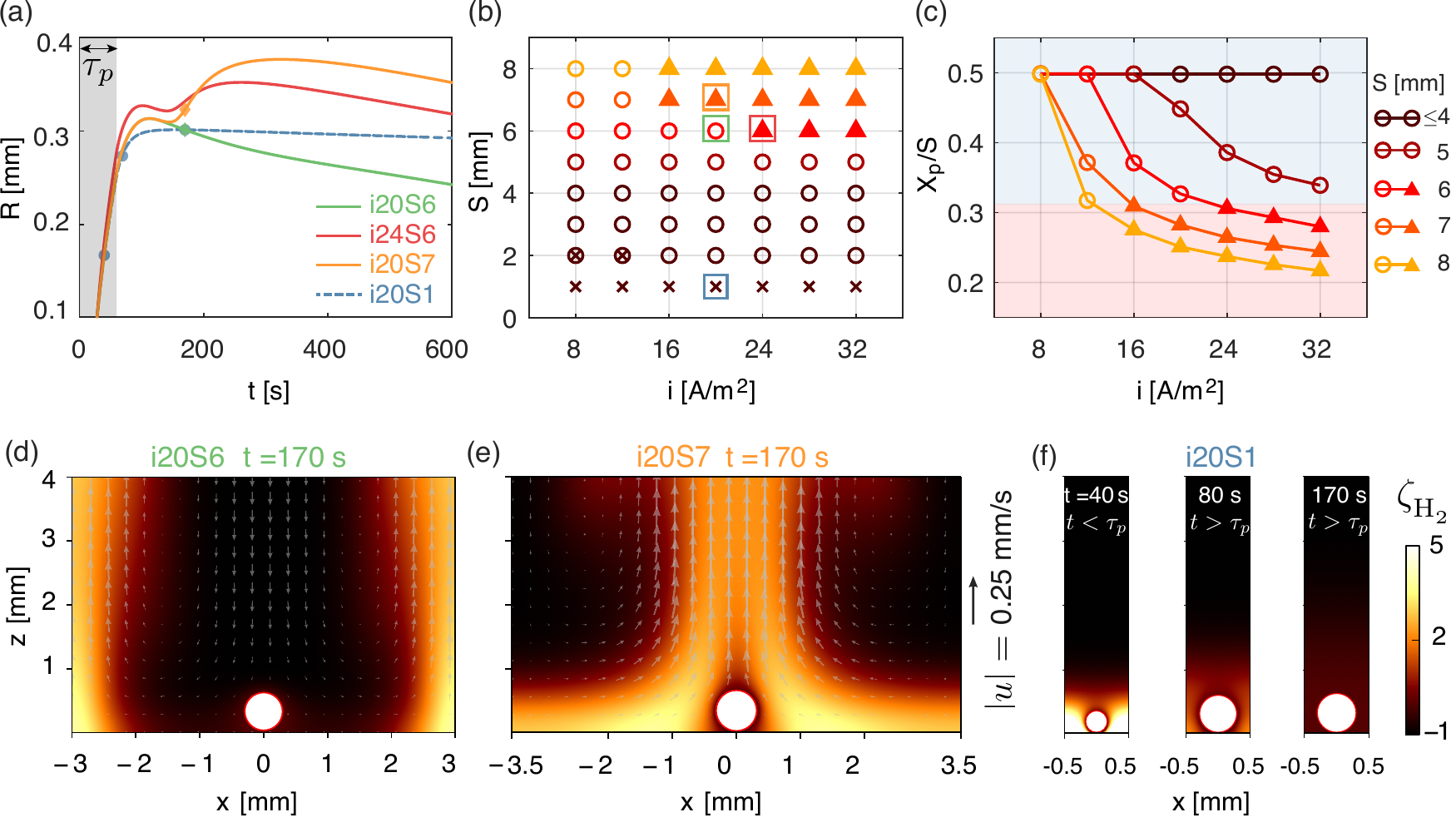}
		\caption{(a) Radius evolution for four parameter  combinations of $i$ and $S$. (b) Full phase map explored for a 60 s pulse with cases shown in (a) marked by squares. Circles (triangles) indicate plumes merging in between (on top of) the bubbles, crosses represent no convection. (Note that the two cases at $S = 2$ mm underwent transitions for continuous currents but not with the 60 s pulse).  (c) Plume detachment location ($x_p/S$) as function of $i$. Symbols as in (b). (d-f) Snapshots of hydrogen contours and velocity vectors corresponding to cases shown in (a): plume between (d) and on top of bubbles (e) and no convection (f). \rev{The reference vector in (e) applies to both panels (d) and (e).} \rev{Full movie is available in the supplementary content.} }
		\label{fig:CurrentSpace}
	\end{figure*}
	
	\subsection{Effect of current density and bubble spacing}  
	In the following, the pulse duration is kept fixed at 60 s as in the experiment, while the current density and box size $S$ are varied systematically. We start the considerations from base case with $i = 20 \ \mathrm{A/m^2}$ and $S = 6$ mm (\rev{$i20S6$}), for which the bubble radius $R(t)$ is shown as a green line in Fig. \ref{fig:CurrentSpace}(a). Even though the parameters  of this case differ from those in Fig. \ref{fig:dissolution}(b), the bubble behaviour appears qualitatively unchanged. However, at a slightly larger box size of $S= 7$ mm (\rev{$i20S7$}, orange line), significant differences arise in the bubble evolution at $t \approx 150$ s, where a secondary growth phase sets in. 
	The reason for this difference is illustrated by the flow patterns in Fig. \ref{fig:CurrentSpace}(d,e). While the plumes rise at the edges of the domain (i.e. halfway between adjacent bubbles) for \rev{$i20S6$} (Fig. \ref{fig:CurrentSpace}(d)), the plumes merge on top of the bubble for \rev{$i20S7$} (Fig. \ref{fig:CurrentSpace}(e)). This implies that at later times, the bubble is no longer surrounded by under-saturated `fresh' electrolyte, but gets exposed to a lateral influx of fluid with high oversaturation $\zeta_{\ce{H2}}$, which leads to the renewed growth phase after the initial dissolution. Given the transient driving, the bubble will also dissolve eventually in this case once the initial boundary layers are drained. Remarkably, also increasing the current from the base case to $i = 24 \ \mathrm{A/m^2}$ (\rev{$i24S6$}) can induce the same phenomenon as shown by the red line in Fig. \ref{fig:CurrentSpace}(a). An overview over the full parameter space in the range $8~\mathrm{A/m^2}\leq i \leq 32~\mathrm{A/m^2}$ and $1~\text{mm} \leq S \leq 8~\text{mm}$ is shown in Fig. \ref{fig:CurrentSpace}(b), where open (full) symbols denote the mode where at later times the plumes merge in between (on top of) the bubbles. From this, it becomes clear that the upward flow is located at the bubble for large $i$ and $S$. This behaviour is related to the lateral density gradient induced by the presence of the bubble: The denser fluid close to the bubble creates a disturbance in the boundary layer (Fig. \ref{fig:DensityContours}) that travels outward and from which eventually the plumes detach. If the disturbance has travelled close enough to, or even reached the boundary at the onset of convection, the plumes will merge there and rise half-way between the bubbles. If, on the other hand, convection sets in while the disturbance is still close to the bubble, the plumes will flap back and merge over the bubble as seen in Fig. \ref{fig:CurrentSpace}(e). Increasing the bubble spacing $S$ increases the distance the disturbance needs to travel before it can interact with the one coming from the adjacent bubble. In contrast, increasing the current density $i$ shortens the time $\tau_c$ before convection occurs and hence also the time during which the disturbance can travel before the plumes detach. 
	
	\begin{figure}[t!]
		\centering 
		\includegraphics[width=0.5\columnwidth]{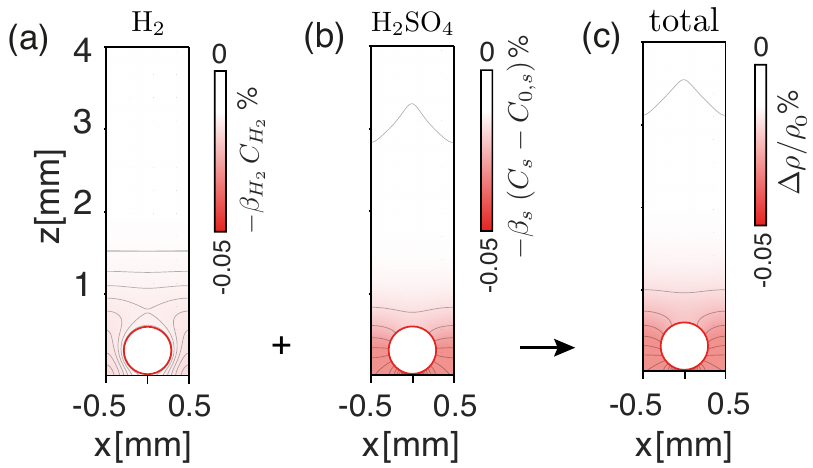}
		\caption{Contribution of local concentration variation of (a) hydrogen and (b) sulfuric acid to the (c) total density fluctuations in the electrolyte solution at $t=80$ s for the case $i20S1$ shown in Figure \ref{fig:CurrentSpace}(f) in the main text.}
		\label{fig:Density-1mm}
	\end{figure}
	
    In order to confirm this picture, we determine $\tau_c$ as the time when the convective transport first equals the diffusive flux. Further, we define the location $x_p$ of the initial plume emission, based on the maximum in the vertical velocity at boundary layer height at time $t= \tau_c$. Details for this procedure are given in \hyperref[sec:AC]{Appendix C}. 
    In Fig. \ref{fig:CurrentSpace}(c), we present the results in the form of $x_p/S$ vs. $i$. These data show that $x_p$ indeed tends to decrease with increasing current density. Most importantly, we also find that the plume location at later times depends on $x_p/S$ as expected from the above argument. In particular, the criterion for the plumes to merge over the bubbles is determined to be $x_p/S\lessapprox 0.31$ from Fig. \ref{fig:CurrentSpace}(c). 

    Finally, when decreasing the bubble spacing drastically to $S=1$ mm (\rev{$i20S1$}), the bubble size is seen to remain approximately constant after the end of the pulse (blue line in Fig. \ref{fig:CurrentSpace}(a)). As shown by the oversaturation contours in Fig. \ref{fig:CurrentSpace}(f), the mass transfer to the bubble effectively balances the production of $\ce{H2}$ in this case. This limits the growth of the hydrogen boundary layer and reduces the buoyancy force. Note that a density difference still arises from the depletion of $\ce{H2SO4}$ (Fig. \ref{fig:Density-1mm}), but the onset of convection is further suppressed by the no-slip condition on the bubble surface, reducing the effective length scale to the bubble spacing instead of the height of the diffusive layer. We therefore observe no convective motion for the cases marked with a cross in Fig. \ref{fig:CurrentSpace}(b), which correspond to low $S$ and low $i$.

    \subsection{The onset of convection}
    
     \begin{figure}[t!]
		\centering
		\includegraphics[width=0.5\columnwidth]{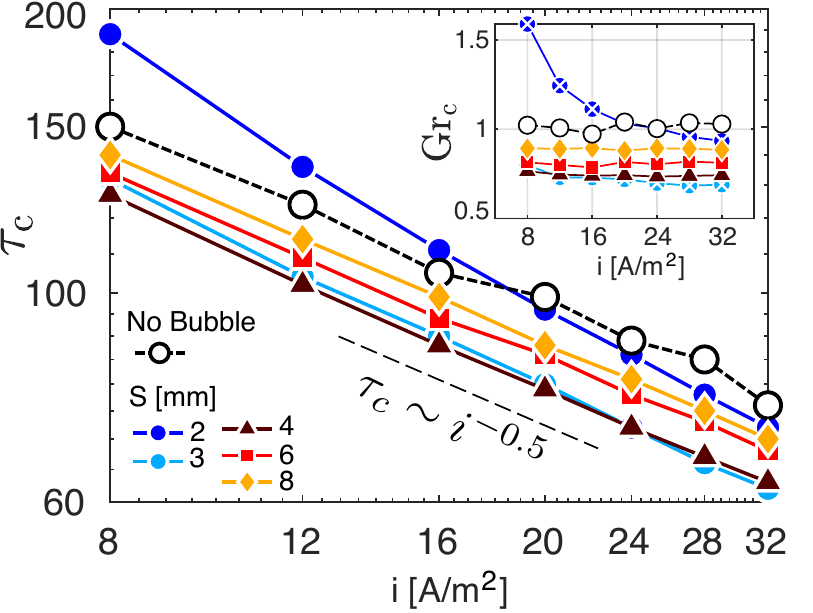}
		\caption{Transition time $\tau_c$ as a function of the current density $i$. The inset shows $Gr_c = Gr(\tau_c)$; note that for $S\leq 2$ mm (marked by a cross), $Gr_c$ was extrapolated from the $Gr(t) = f(i,t)$ curve obtained at larger $S$.}
		\label{fig:Gr}
	\end{figure}

    Next, we will examine the onset of convection and study how this is influenced by the presence of the bubbles. In order to render the considerations independent of the pulse duration $\tau_p$, a continuous current is applied in the simulations for this purpose.
    In Fig. \ref{fig:Gr}, we present results for the time of convection onset $\tau_c$ for different bubble spacings $S$ as a function of $i$. In addition, the plot also contains data for a reference case without bubbles. Initially focusing on $S\geq 4$ mm for which a largely undisturbed region exists in between the bubbles, $\tau_c$ is seen to decrease with $i$ according to roughly $\tau_c \sim i^{-1/2}$. Moreover, $\tau_c$  at constant $i$ is largest for the case without bubbles and decreases as the bubble spacing $S$ is reduced. To gain a better insight into these trends, we define a Grashof number
    \begin{equation}
    	\mathrm{Gr}=\frac{g\delta^3}{\nu^2}\frac{-\Delta \rho (z =0)}{\rho_0},
    	\label{eq:Gr}
    \end{equation}
    which compares buoyancy with viscous forces. Here, the height $\delta$ of the initial diffusion boundary layer is defined based on the instantaneous density profile normal to the electrode (see \hyperref[sec:AD]{Appendix D}). Eq. \eqref{eq:Gr} therefore encompasses the full density difference, which originates to approximately equal parts from the distributions of $\ce{H2}$ and $\ce{H2SO4}$ (see Fig. \ref{fig:DensityContours} and \ref{fig:DiffusionDepth} in \hyperref[sec:AC]{Supporting Infromation}). The Grashof number is closely related to the Rayleigh number, which is also frequently used in this context \cite{Tan1996,Pande2021,Karatay2016,DeValenca2017,Tassel2007}. The use of $Gr$ is preferred here since its definition is independent of the mass diffusivities, which differ for $\ce{H2}$ and $\ce{H2SO4}$. Generally speaking, $Gr$ is an increasing function of time as both $\delta$ and $\Delta \rho$ increase with $t$. In the inset of Fig. \ref{fig:Gr}, we have plotted $Gr_c(t = \tau_c)$ at the onset of convection. For $S\geq 4$mm, the value of $Gr_c$ is found to be independent of the current density $i$. Still, the value of the critical Grashof number beyond which convection sets in, $Gr_c$, depends on the precise bubble configuration and decreases from $Gr_c\approx 1$ in the absence of bubbles\footnote{Using the Schmidt number $Sc =404$ of \ce{H2SO4}, this is consistent with the range of critical Rayleigh numbers  $ 320 \le Ra_c = Gr Sc \le  817$ reported for temperature \cite{Sparrow1964} and gas diffusion \cite{Tan1992} boundary layers.} to $Gr_c \approx 0.75$ for $S = 4$ mm. This gives evidence that the presence of the bubbles destabilizes the boundary layer such that buoyancy driven convective motion sets in earlier. Having established that $Gr_c = const. $ for large enough bubble spacings, we can also explain the scaling of $\tau_c$: \rev{From the solution of a constant flux diffusion problem} \cite{Bejan1993}, \rev{we get the scalings $-\Delta \rho (z =0)\sim it^{1/2}$  and $\delta \sim t^{1/2}$, such that the Grashof number grows according to $Gr \sim it^2$. The latter results in $t_c \sim i^{-1/2}$, exactly as observed in Fig.} \ref{fig:Gr}.
    
    When decreasing the bubble spacing below $S= 4$ mm, we notice that $\tau_c$ does not decrease further at $S=3$ mm and eventually increases again for $S= 2$ mm. Again, this is a combined effect of the $\ce{H2}$ transfer into the bubbles and suppression of flow by their presence. At lower $i$, the longer transition times render the mass transfer into the bubble more relevant, which leads to a deviation from the $\tau_c\sim i^{-1/2}$ scaling, especially at $S= 2$ mm.
    The same mechanism is also reflected in a significant increase of $Gr_c$ with decreasing $i$ in the inset for $S=3$ mm and even more prominently for $S= 2$ mm. No convection was observed for the tightest spacing of $S = 1$ mm even with continuous driving.

    \subsection{Effect of bubble clustering}
    The results so far present convincing evidence and insight into the role of convection in the evolution of the hydrogen bubbles on the electrode surface. Yet, single bubble simulations fail to reproduce the experimental results quantitatively (see Fig. \ref{fig:dissolution}(b)). Further, these results also did not feature the change in dissolution rate, which is evident to varying degrees for all of the experimental recordings in Fig. \ref{fig:dissolution}(a) at about 200 s after the end of the pulse. In the following, we will demonstrate that collective effects of multiple interacting bubbles can explain these differences.

    For this purpose, we consider the $3\times3$ cluster of bubbles as shown in Fig. \ref{fig:setup}(b). For all simulations with  clusters, the box size is fixed to $S= 4$ mm (in all three directions) and the experimentally measured current density during the 60 s pulse is used (see Fig. \ref{fig:ExpSetup} (c)). Thus, the only parameter which is varied is the inter-bubble spacing $S_c$. 
	
\begin{figure*}[t!]
		\centering
		\includegraphics[width=1\columnwidth]{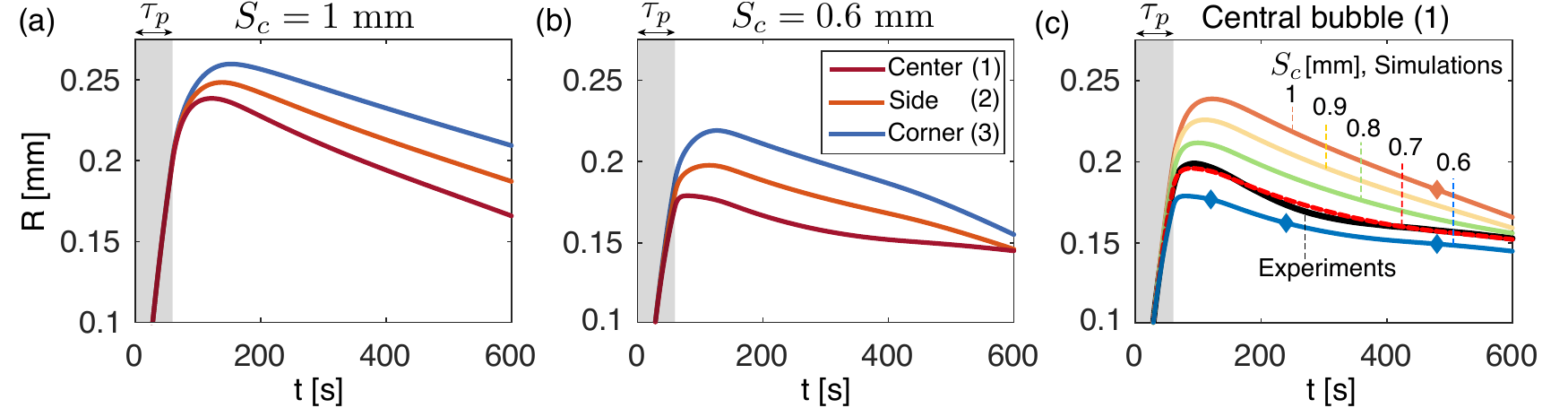}
		\caption{Bubble radius evolution in the cluster configuration (a) with $S_c = 1$ mm and (b) $S_c = 0.6$ mm. (c) $R(t)$ for the central bubble only at varying $S_c$.}
		\label{fig:SimSpacing}
	\end{figure*}

    The time traces of $R(t)$ in Fig. \ref{fig:SimSpacing}(a) display a behaviour that is consistent with the convective pattern of plumes rising in between bubbles observed earlier. As expected, there is no difference in the size of  bubbles at different locations during the growth period. However, such differences do arise during the dissolution stage, where the central bubble starts dissolving the earliest and at the fastest rate. The transition from growth to dissolution (and to a lesser extent also the final dissolution rate) are progressively slower for the bubbles at the sides and in the corners. This overall picture continues to apply also if the cluster spacing is reduced to $S_c = 0.6$ mm in Fig. \ref{fig:SimSpacing}(b). The decreased spacing does, however, lead to a fast onset of dissolution for all bubbles. Moreover, the evolution of the bubble radius with time now also features the distinct change in slope at around $t=300$ s, similar to the experimental observations.
    
\begin{figure*}[t!]
		\centering
		\includegraphics[width=1\columnwidth]{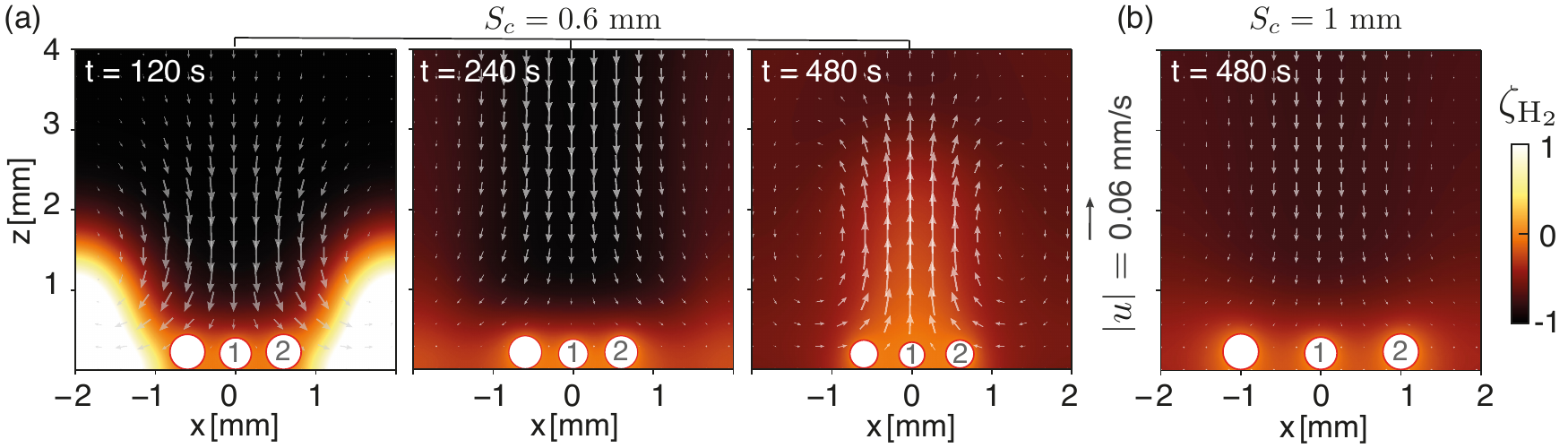}
		\caption{Snapshots of the \ce{H2} oversaturation along with velocity vectors for (a) $S_c = 0.6$ mm and (b) $S_c = 1$ mm. \rev{The reference vector applies to all panels.} \rev{Full movie is available in the supplementary content.}}
		\label{fig:Shielding}
	\end{figure*}

\begin{figure}[t!]
		\centering 
		\includegraphics[width=0.5\columnwidth]{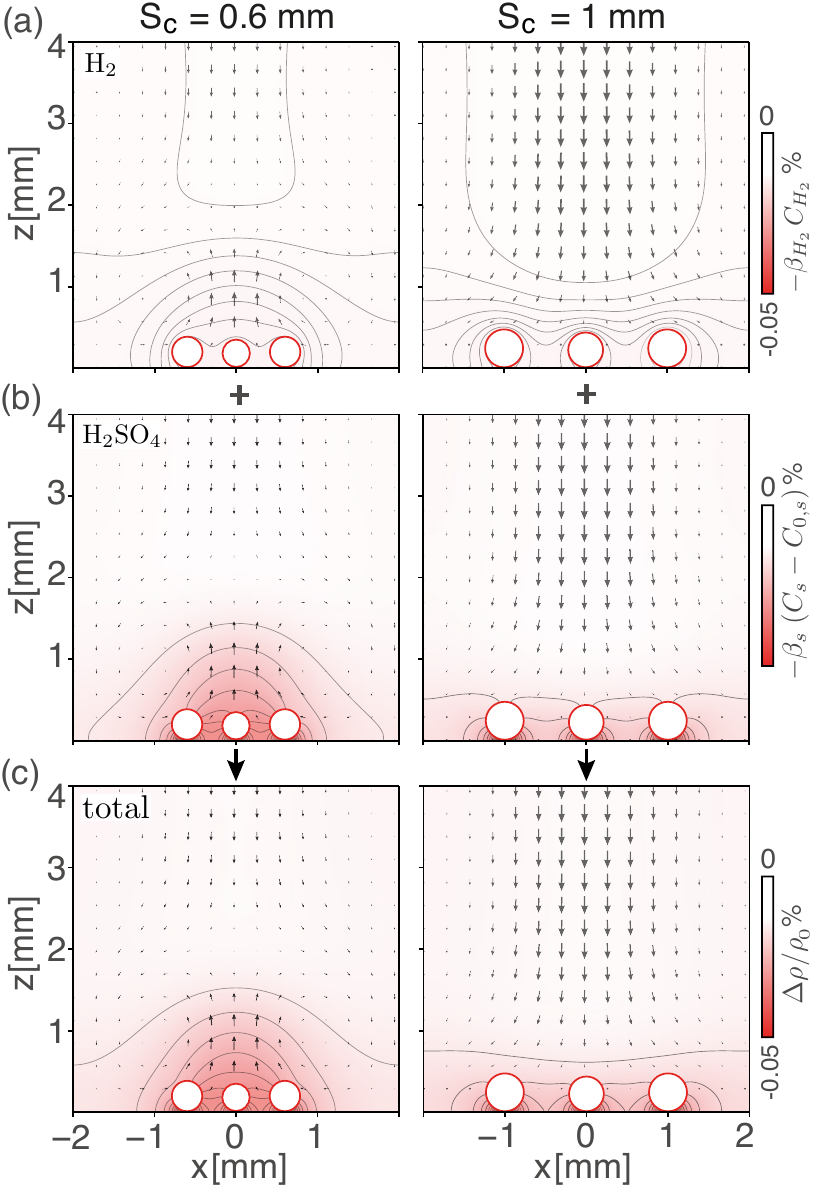}
		\caption{Contribution of local concentration variation of (a) hydrogen and (b) sulfuric acid to the (c) total density fluctuations in the electrolyte solution at $t=400$ s. The current density is taken from the experimentally measured values (black curve in Figure \ref{fig:ExpSetup}(c)). The distance between the bubbles in the network is  $S_c=0.6$ mm for left panels and $S_c=1$ mm for right panels.}
		\label{fig:DenMultiple}
	\end{figure}

     Contours plots of the hydrogen oversaturation $\zeta_{\ce{H2}}$ along with the convective patterns in Fig. \ref{fig:Shielding}(a) help explain these findings. Since the plumes rise in between the clusters, the downward flow is consequently centered on the bubble in the middle (bubble 1 in Fig. 9), which is therefore most exposed to the undersaturated electrolyte compared to those further out (bubbles 2 and 3). This behaviour is similar for $S_c = 0.6$ mm and $S_c = 1$ mm. There are significant differences however at later times. At $t = 480$ s, an upward flow forms over the dissolving bubble cluster with $S_c = 0.6$ mm, whereas such a pattern is entirely absent in the case with $S_c = 1$ mm in Fig. \ref{fig:Shielding}(b). An analysis of the corresponding density contours (Fig. \ref{fig:DenMultiple}) reveals that the upward flow is not predominantly driven by variations in the $\ce{H2}$ field resulting from the bubble dissolution. \rev{A decisive factor is rather that the depletion of $\ce{H2SO4}$ caused by the reaction cannot be `washed out' effectively due to the blockage by the tightly spaced bubbles. In this way, lower density electrolyte persists within the cluster and helps drive the observed upward convection at late times.} Once convection sets in, the well-known shielding effect \cite{Carrier2016,Laghezza2016,Chong2020} reduces the dissolution rate of central bubble, while slightly increasing the dissolution rate of the other bubbles (compare also Fig. \ref{fig:SimSpacing}(b) at later times).
     
    The dependence of the general size of the central bubble on $S_c$ is considerable, as the data in Fig. \ref{fig:SimSpacing}(c) prove. An excellent match between the experimental data and our modeling results is obtained for $S_c = 0.7$ mm, which is indeed very close to the distance to the neighbouring bubble observed in Fig. \ref{fig:ExpSetup}(b). It therefore appears very likely that collective effects due to the inhomogeneous bubble distribution play an important role in the experiment. This remains true, even if unaccounted effects, such as the presence of dissolved air, may alter the $R(t)$ curves slightly.
    
	\section{Conclusion}
	Our combined experimental and numerical analysis firmly established the relevance of solutal convection for  bubble evolution during water electrolysis. The experimentally observed bubble behaviour was shown to be inconsistent with pure diffusive transport, while experiments and simulations were in excellent agreement when natural convection due to buoyancy effects was considered.
	While appropriate for micro-electrodes \cite{Yang2018,Massing2019,Meulenbroek2021}, our results suggest that convective effects cannot be neglected when larger electrodes are considered \cite{VanDerLinde2017,VanDerLinde2018,Hossain2020}. For example, estimating based on the $\ce{H2}$ concentration only, a critical value for the onset of convection of $Gr \approx 1$ should be reached after about 100s for the conditions  reported in \citeauthor{VanDerLinde2017} \cite{VanDerLinde2017}, while their experiments lasted for hours. Our results further show that the presence of bubbles can decrease the stability threshold of the diffusive boundary layers, rendering the system even more prone to convective effects. For the present conditions, this destabilization occurs if $S\geq 4$ mm, while the onset of convection is delayed or even suppressed entirely if the distance between bubbles is smaller than $S \leq 1$ mm. We further demonstrated that convective patterns and especially their impact on the bubble evolution vary significantly, depending on the design parameters. This may open up avenues to control flow features to achieve a desired bubble behaviour by providing nucleation sites with optimized spacings. However, there still remain open questions. These pertain e.g. to the potential effect of spatially varying current density due to the presence of the bubbles \cite{Dukovic1987}. Further, a more complete treatment of the problem especially at high values of $i$ and for tight bubble spacing should also include \rev{supporting electrolyte and} the effect of Marangoni convection \cite{Lohse2020,Hossain2020}. Finally, allowing for bubble detachment in the simulations will enable accessing stages after the initial transient.\\
	
\section*{Acknowledgements}
	This work was supported by the Netherlands Center for Multiscale Catalytic Energy Conversion (MCEC), an NWO Gravitation programme funded by the Ministry of Education, Culture and Science of the government of the Netherlands.  This research also received funding from The Netherlands Organization for Scientific Research (NWO) in the framework of the fund New Chemical Innovations, project ELECTROGAS (731.015.204), with financial support of Akzo Nobel Chemicals, Shell Global Solutions, Magneto Special Anodes (an Evoqua Brand), and Elson Technologies.  This project also received funding from the European Union’s Horizon 2020 research and innovation programme (grant agreement No. 950111 BU-PACT, No. 740479 DDD, and the Marie Skłodowska‐Curie grant agreement No 801359). We also acknowledge PRACE for awarding access to MareNostrum at Barcelona Supercomputing Center (BSC), Spain (Project 2020225335 and 2020235589) and the Max Planck Center Twente for Complex Fluid Dynamics for financial support.

\appendix
\section{Electrolyte transport equation}
\label{sec:AA}
\renewcommand{\theequation}{A.\arabic{equation}}
\setcounter{equation}{0}

	Here the derivation of the advection-diffusion equation for \ce{H2SO4} (j=s in Eq. \eqref{salt}) will be presented. We start from the mass-transport equations for dissolved ions given by
	
	\begin{align}\label{cation}
		\frac{\partial C_1}{\partial t} + \left ( \textbf u \cdot \bm \nabla \right ) C_1 =  D_1 \bm \nabla^2 C_1 + D_1 z_1 \frac{ F}{RT} \bm  \nabla \cdot \left(  C_1 \bm \nabla \phi \right),
	\end{align}
	and
	\begin{align}\label{anion}
		\frac{\partial C_2}{\partial t} + \left ( \textbf u \cdot \bm \nabla \right ) C_2 =  D_2 \bm \nabla^2 C_2 + D_2  z_2 \frac{ F}{RT} \bm  \nabla \cdot\left(  C_2 \bm \nabla \phi \right),
	\end{align}

	\noindent where subscripts 1 and 2 denote \ce{H+} and \ce{SO4^2-} ions, respectively, $\phi$ refers to the electric potential field and $z_k$ denotes the ionic valence i.e., $z_1=+1$ and $z_2=-2$.  Employing the electroneutrality condition
	
	\begin{equation}\label{EN}
		z_1 C_1 = -z_2 C_2,
	\end{equation}
	
	\noindent equation \eqref{anion} can be expressed in terms of $C_1$ as 
	
	\begin{align}\label{anion2}
		-\frac{z_1}{z_2} \frac{\partial C_1}{\partial t}  - \frac{z_1}{z_2}  \left ( \textbf u \cdot \bm \nabla \right ) C_1  = -  \frac{z_1}{z_2}  D_2 \bm \nabla^2 C_1	- D_2  z_1 \frac{ F}{RT} \bm  \nabla \cdot \left(  C_1 \bm \nabla \phi \right).
	\end{align}
	
	\noindent \noindent Multiplying equation \eqref{anion2} by $D_1$ and subtracting it form equation \eqref{cation} multiplied by $D_2$ gives
	\begin{align}\label{salt1}
		\left( D_2 - D_1 \frac{z_1}{z_2}\right) \frac{\partial C_1}{\partial t} + \left(D_2 -D_1 \frac{z_1}{z_2}  \right) \left ( \textbf u \cdot \bm \nabla \right ) C_1 =  D_1 D_2 \left( 1 - \frac{z_1}{z_2} \right)   \bm \nabla^2 C_1.
	\end{align}
	
	\noindent Rearrangement of the terms in equation \eqref{salt1} by taking into account that $C_{\ce{H2SO4}} = C_{\ce{H+}}/2$ (according to the electroneutrality condition and full dissociation of sulfuric acid in water) yields 
	\begin{equation}\label{salt2}
		\frac{\partial C_s}{\partial t} +  \left ( \textbf u \cdot \bm \nabla \right ) C_s =  D_s  \bm \nabla^2 C_s ,
	\end{equation}
	
	\noindent where the electrolyte diffusivity $D_s$ is defined as
	
	\begin{equation}\label{SaltDiff}
		D_{\ce{s}} = \frac{D_1 D_2  \left( z_1 - z_2 \right)}{z_1 D_1 - z_2 D_2}.
	\end{equation}

	\noindent Accordingly,  equations \eqref{cation} and \eqref{anion} are simplified to the single equation \eqref{salt2} thereby eliminating the migration terms.\\
	
	The proton is reduced at the electrode surface. Using the same steps as above for Eq. \eqref{salt2}, the associated flux of \ce{H+} at the boundary can be related to the current density by
	
	\begin{equation}\label{BC1}
		\frac{i}{(n_e / s_1)F} = D_1 \left( \frac{\partial C_1 }{\partial z} + z_1 C_1 \frac{F}{RT} \frac{\partial \phi}{\partial z} \right)_{z=0}.
	\end{equation}
	
	\noindent Since the anion is not consumed in the electrochemical reaction on the electrode surface, its flux is zero there. Thus, we obtain
	
	\begin{equation}
		\left( \frac{\partial C_2}{\partial z} \right) _{z=0} = -z_2 C_2 \frac{F}{RT}\left( \frac{\partial \phi}{\partial z} \right)_{z=0} ,
	\end{equation}
	
	\noindent which along with electro-neutrality condition yields 
	
	\begin{equation}\label{BC2}
		\left( 	\frac{\partial C_2}{\partial z}\right) _{z=0} = 	-\frac{z_1}{z_2} \left( \frac{\partial C_1}{\partial z} \right) _{z=0} = z_1 C_1 \frac{F}{RT}\left( \frac{\partial \phi}{\partial z}, \right)_{z=0} .
	\end{equation}
	
	\noindent Again taking into account that $C_{\ce{H2SO4}} =C_{\ce{H+}}/2$, equation \eqref{BC2} is used to eliminate the migration terms in  \eqref{BC1} according to
	
	\begin{equation}
		\frac{i}{(n_e / s_1)F} = 2D_1 \left(1 - \frac{z_1}{z_2} \right) \left( \frac{\partial C_s}{\partial z} \right) _{z=0},
	\end{equation}
	
	\noindent which is used as boundary condition for equation \eqref{salt2}.\\
	
	\noindent Again taking into account that $C_{\ce{H2SO4}} =C_{\ce{H+}}/2$, equation \eqref{BC2} is used to eliminate the migration terms in  \eqref{BC1} according to
	
	\begin{equation}
		\frac{i}{(n_e / s_1)F} = 2D_1 \left(1 - \frac{z_1}{z_2} \right) \left( \frac{\partial C_s}{\partial z} \right) _{z=0},
	\end{equation}
	
	\noindent which is used as boundary condition for equation \eqref{salt2}.\\
	
	\section{Numerical methods}\label{sec:AB}
	
	Direct numerical simulations are used to solve the system of equations \eqref{NS} and \eqref{cont} in a three dimensional Cartesian domain as depicted in Fig. \ref{fig:setup} in the main text. Spatial terms are discretized using a second-order accurate finite difference method on a staggered grid. A fractional-step third-order Runge-Kutta scheme, in combination with a Crank-Nicolson scheme for the viscous terms are employed to perform the time marching \cite{VanderPoel2015,R1996}. Periodic boundary conditions for the velocity components and scalar fields are employed at side walls of the Cartesian domain in wall-parallel directions. An outflow boundary condition is applied at the top boundary, through which the diffusive and advective fluxes of both velocity and scalar fields are conserved. The solver is coupled with a versatile moving least squares (MLS) based immersed boundary method (IBM), \cite{DeTullio2016,Spandan2017} which uses a triangulated grid network called Lagrangian markers (Fig. \ref{fig:setup}(a)) to enforce the gas-liquid interfacial boundary conditions, including saturation concentration for hydrogen and no-flux for other species alongside no-slip and no-penetration conditions for velocity field, and transfer these quantities back to the underlying Eulerian mesh. Therefore, any flow field generated inside the bubble is disregarded as it is irrelevant to the flow physics outside the bubble. The no-slip boundary condition on the bubble is chosen in order to represents a fully contaminated bubble surface \cite{Takagi2011}.
	
	Finally, the location of Lagrangian markers is updated in time based on equation \eqref{radius}. It is further worth mentioning that the concentration gradient $\left(\bm{\nabla} C_j \cdot \hat{\textbf n}\right) \big |_\Sigma$ at the bubble interface is calculated through extending a probe normal to the barycentre of each triangulated Lagrangian face and determining the scalar concentration at the tip of the probe by an additional MLS interpolation. 
	
	The computational domain has a fixed height of 4 mm in all cases and has a quadratic outline in the horizontal (parallel to the electrode) plane with varying side length $S$. The initial bubble size is limited by resolution requirements. Here, we have chosen the initial diameter of the bubble to be 1/20 of the domain height and used $\approx13$ grid points to resolve the initial bubble diameter after checking grid independence. This choice offered a reasonable compromise between starting with the smallest bubble possible and keeping the computational cost at bay. 
	The time at which the bubble is initialized in the simulations (here $28.21$ \text{s}) with diameter of $0.2 \ \text{mm}$ has been chosen to match the experimental data (black curve in Fig. \ref{fig:ExpSetup}(c)). The initalization time was also kept constant when varying the current density from the experimental value for consistency. We ran tests with an earlier bubble injection at higher currents in order to confirm that the choice of the bubble initialization time did not change our results significantly. 
	
	Physical properties of the analyzed electrochemical system are tabulated in table \ref{table1}. The molar expansion coefficient of hydrogen in sulfuric acid varies depending on the initial concentration of sulfuric acid in water and we have computed it using the correlation proposed by \citeauthor{Vogt1992} \cite{Vogt1992}. The full set of numerical parameters is listed in table \ref{table2}.
	
	\begin{table}[h!]
	\def\arraystretch{1.2}
		\centering	
		\caption{Physical properties of the analyzed system. $k_{\ce{H2}}$ is Henry's constant such that $C_{\ce{H2},sat}=k_{\ce{H2}}P_0$.}
		\label{table1}
		\begin{tabular}{ l l }
			\hline
			Properties & Unit\\
			\hline
			$\left( C_\mathrm{H_{2}SO_4} \right) _0 \ = \ 100$ \ & $\mathrm{mol \ m^{-3}}$\\
			$T_\infty \ = \ 298$ \ & $\mathrm{K}$ \\
			$P_0 \ = \ 1$ \ & $\mathrm{bar}$ \\ 
			$\rho_L \ = \ 1030 $ \ & $\mathrm{kg \ m^{-3}}$ \\ 
			$\nu_L \ = \ 0.94 \times10^{-6}$  \ &  $\mathrm{m^2 \ s^{-1}}$ \\ 
			$D_{\ce{H+}} \ = \ 9.308 \times10^{-9}$  \ &  $\mathrm{m^2 \ s^{-1}}$ \\ 
			$D_{\ce{SO4^2-}} \ = \ 1 \times10^{-9}$  \ &  $\mathrm{m^2 \ s^{-1}}$ \\ 
			$D_{\ce{H2}} \ = \ 3.7 \times10^{-9}$  \  & $\mathrm{m^2 \ s^{-1}}$ \\ 
			$k_{\ce{H,H2}} \ = 7.2 \times 10^{-6}$  \   &    $\mathrm{mol \ m^{-3} \ Pa^{-1}}$				\\
			$\beta_{\ce{H_2}} \ = \ +11.5 \times10^{-6}$  \  & $\mathrm{m^3 \ mol^{-1}}$ \\ 
			$\beta_{\ce{H_2SO_4}} \ = \ -62 \times10^{-6}$  \ &  $\mathrm{m^3 \ mol^{-1}}$ \\ 
			\hline
		\end{tabular}
	\end{table}
	
		\begin{table}[h!]
		\def\arraystretch{1.2}
		\centering	
		\caption{Numerical setup information}
		\label{table2}
		\begin{tabular}{ l l l }
			\hline
			Parameter & Value & Unit\\
			\hline
			Domain size \ & $4 \times S \times S$ & $\mathrm{mm}$\\
			Initial bubble diameter \ & 0.2 & $\mathrm{mm}$ \\
			Grid No. per initial\\
		    bubble diameter\  & $13$ \\ 
			Time step\ &  $0.005-0.05$ \ & $\mathrm{s}$ \\ 
			Bubble injection time\  & $28.21$ \ & $\mathrm{s}$\\
			\hline
 		\end{tabular}
    	\end{table}
    	
    \section{Transition time and gas plumes location}\label{sec:AC}
    \renewcommand{\theequation}{C.\arabic{equation}}
    \setcounter{equation}{0}
    
\begin{figure}[t!]
		\centering
		\includegraphics[width=0.75\columnwidth]{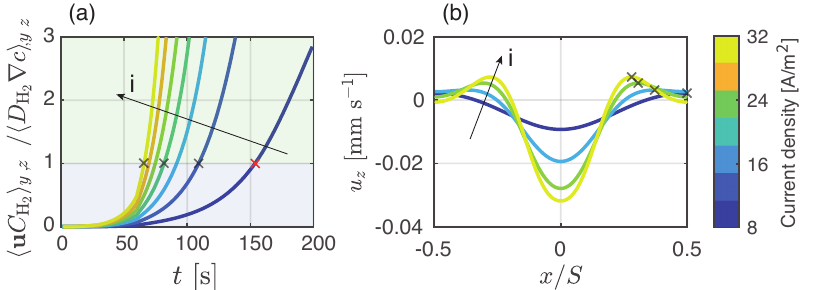}
		\caption{(a) Ratio of advective to diffusive fluxes of hydrogen at $S = 6mm$. Different linis represent varying current density in the range from 8 $\mathrm{A/m^2}$ to 32 $\mathrm{A/m^2}$. Cross markers indicate the transition time measured at the instants when the ratio of the fluxes is unity. (b) horizontal profile of the vertical component of the velocity ($u_z$) at the edge of the hydrogen boundary layer ($\delta_\mathrm{H_2}$) at transition times ($\tau_c$) obtained from panel (a). Cross markers locate the peaks in the profile based on which $x_p$ is determined. Current density is varied from 8 $\mathrm{A/m^2}$ to 32 $\mathrm{A/m^2}$. }
		\label{fig:tauc_vel}
	\end{figure}

	We base the criterion for the onset of convection on the $\ce{H2}$ distribution and define the transition time $\tau_c$ as the time at which the averaged advective flux first exceeds the diffusive transport, i.e.,
		\begin{equation}
			\langle \textbf{u}C_{\mathrm{H_2}} \rangle_{y,z} \ge \langle D_{\mathrm{H_2}} \nabla C_{\mathrm{H_2}} \rangle_{y,z},
		\end{equation} 
	where $\langle \rangle_{y,z}$ denotes an average over the midplane of the domain.
	Fig. (\ref{fig:tauc_vel}) displays samples of the ratio of the advective to diffusive fluxes for $S=6$ mm at different current densities, where $\tau_c$ is marked with crosses. 
	
	We used the location of the gas plumes at transition time to distinguish two different modes of the convective pattern, which can lead to either enhanced growth or dissolution of the bubble. 
	To determine the plume detachment position $x_p$, we consider the horizontal profile of the vertical velocity ($u_z$) at $z=\delta_\mathrm{H_2}$ as shown in Fig. \ref{fig:tauc_vel}(b), where  $\delta_\mathrm{H_2}$ is the hydrogen boundary layer thickness sufficiently far from the bubble. We then define $x_p$ as the location of the  peaks in the velocity profile as indicated Fig. \ref{fig:tauc_vel}(b).

	\section{Effective diffusion depth}
    \renewcommand{\theequation}{D.\arabic{equation}}
    \setcounter{equation}{0}
    
\begin{figure}[t!]
		\centering 
		\includegraphics[width=0.5\columnwidth]{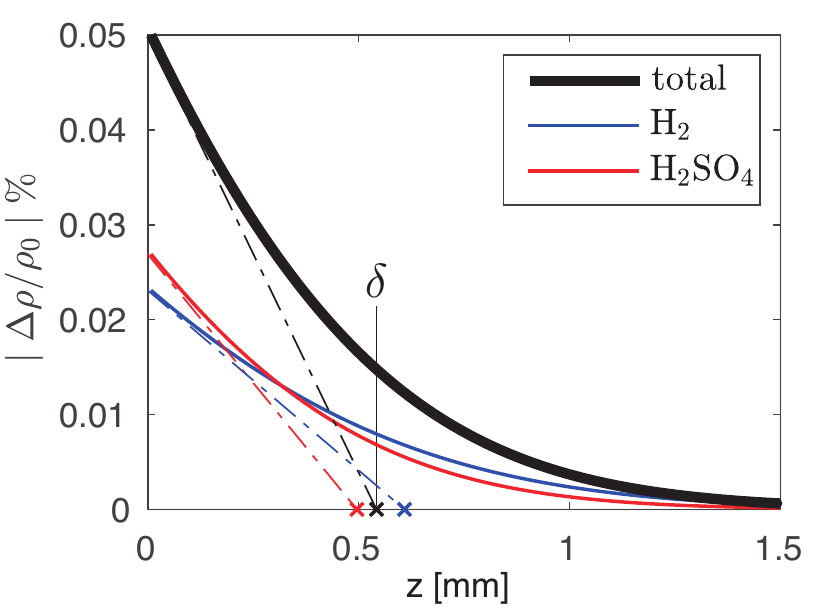}
		\caption{Total density variation profile and contributing components for a simulation without bubble and $i=24$ $\mathrm {A/m^2}$. Profiles are plotted in the centerline of the mid-plane ($x/S=0$) at $t=80$ s. Dashed lines indicate the linear fit at the electrode surface ($z=0$) to each profile and crosses mark the corresponding locations of $\delta$.}
		\label{fig:DiffusionDepth}
	\end{figure}

	Here, we explain the approach employed for measuring the instantaneous effective diffusion depth $\delta$, which accounts for the density variations resulting from the change in concentration of $\ce{H2SO4}$ and hydrogen gas adjacent to the electrode. A typical density profile and its constituents at $t=80$ s are plotted in Fig. \ref{fig:DiffusionDepth}. As the Fig. shows, both hydrogen enrichment and electrolyte depletion contribute approximately equally to the total density variation. We define $\delta$ as used in the definition of $Gr$ in Eq. \ref{eq:Gr} based on the total density profile according to 
	\begin{equation}
	    \delta = \frac{\Delta \rho}{ \partial_z(\Delta \rho)}\vert_{z=0}.
	\end{equation}
	This value is indicated by a black marker in Fig. \ref{fig:DiffusionDepth}.
	 The ratio of the diffusivities  for hydrogen and the sulfuric acid is $\sqrt{D_\mathrm{H_2}/D_s}\approx 1.22$, such that the effective diffusion depths based on these profiles (also included in the figure) differ slightly. 
	
\bibliographystyle{elsarticle-num-names}
\newpage
\bibliography{cas-refs}





\end{document}